# Integrating Attentional Factors and Spacing in Logistic Knowledge Tracing Models to Explore the Impact of Training Sequences on Category Learning

Meng Cao
University of Memphis
mcao@memphis.edu

Philip I. Pavlik Jr.
University of Memphis
ppavlik@memphis.edu

Wei Chu
University of Memphis
wei.chu@memphis.edu

Liang Zhang
University of Memphis
lzhang13@memphis.edu

## ABSTRACT
In category learning, a growing body of literature has increasingly focused on exploring the impacts of interleaving in contrast to blocking. The sequential attention hypothesis posits that interleaving draws attention to the differences between categories while blocking directs attention toward similarities within categories [4, 5]. Although a recent study underscores the joint influence of memory and attentional factors on sequencing effects [31], there remains a scarcity of effective computational models integrating both attentional and memory considerations to comprehensively understand the effect of training sequence on students' performance. This study introduces a novel integration of attentional factors and spacing into the logistic knowledge tracing (LKT) models [22] to monitor students' performance across different training sequences (interleaving and blocking). Attentional factors were incorporated by recording the counts of comparisons between adjacent trials, considering whether they belong to the same or different category. Several features were employed to account for temporal spacing. We used crossvalidations to test the model fit and predictions on the learning session and posttest. Our findings reveal that incorporating both attentional factors and spacing features in the Additive Factors Model (AFM) significantly enhances its capacity to capture the effects of interleaving and blocking and demonstrates superior predictive accuracy for students' learning outcomes. By bridging the gap between attentional factors and memory processes, our computational approach offers a more comprehensive framework for understanding and predicting category learning outcomes in educational settings.

## 1. INTRODUCTION
How to present examples to optimize category learning and generalization is crucial in inductive category learning [3, 9, 17, 18, 19, 26, 28]. The sequence of these exemplars significantly influences what we attend to, and how we encode and memorize new categories [5]. While extensive research has been dedicated to identifying effective sequencing methods in category learning, a growing body of literature has focused on exploring the impacts of two different sequencing methods: interleaving and blocking.

Interleaving means presenting one exemplar from one category followed by an exemplar from another category (e.g., $A_1B_1C_1A_2B_2C_2A_3B_3C_3$, where A, B, and C represent different categories and 1, 2, and 3 denote different exemplars). In contrast, blocking entails grouping all exemplars from the same category in a single block so that learners learn one category at a time (e.g., $A_1A_2A_3B_1B_2B_3C_1C_2C_3$). The benefit of interleaving over blocking has been found many times in empirical research on learning visual materials like paintings, naturalistic photos, and artificial images [12, 15, 32]. However, under specific circumstances, blocking can be more beneficial compared to interleaving [2, 3, 4, 5, 10, 11, 30, 32]. For example, Carvalho and Goldstone [3] have found that low-similarity stimuli were more effectively learned when presented in blocked sequences.

Researchers have proposed several theoretical accounts of the interleaving and blocking effects [4, 5, 12, 14]. Yan and colleagues [31] have integrated these diverse theoretical accounts into a two-stage framework. Remarkably, they provided empirical support for this framework using a meta-analytic approach. Their findings indicated that a combination of memory and attentional factors accounts for a significantly greater proportion of variance in sequencing effects compared to attentional factors in isolation. Within this framework, sequencing effects play a pivotal role in shaping category learning through two distinct sequential phases. Firstly, in the attention-based stage, learners are tasked with directing their focus towards the pertinent features of the category structure and recognizing its similarities. At this stage, interleaving encourages discrimination between categories, while blocking promotes the formation of commonality abstractions within categories. Secondly, in the memory-based stage, learners must commit to memory the cluster of features and establish associations between these features and category labels. When practice is spaced over time instead of studied in a single session, forgetting is also decelerated [21]. Interleaving naturally spaces exemplars from the same category apart and the spacing interval is determined by the number of intervening categories. For instance, in a sequence like ABCDABCD, the intervals between the two 'A's are spaced by 3 items. Conversely, in blocking, exemplars from the same category are presented consecutively without any temporal lag or spacing between them. Therefore, one significant drawback of blocking becomes apparent: it lacks spacing.

Recognizing the significance of both attentional factors and spacing in comprehending the effects of training sequences (interleaving and blocking), it is essential for computational models to integrate both aspects simultaneously. While models have been developed to address attentional factors [1, 6] and others track memory in relation to spacing [8, 23, 29], none of these explicitly incorporate both factors concurrently [31]. For example, the Sequential Attention Theory Model (SAT-M) [6] focuses solely on attentional factors by incorporating the encoding strength of features, which depends on the sequence of the study, into the Generalized Context Model (GCM) [20]. The GCM is a special case of exemplar models where learners need to make a categorical selection based on the similarities between the stimulus and all the stored exemplars [7, 16, 24]. The model assumes a person memorizes all prior instances without forgetting and lacks components addressing the spacing effect. Additionally, a recent PFA-Categorization model addresses attentional factors by incorporating the prototype model, where learners make a categorical selection based on the similarities between the stimulus and the prototype in memory, into the Performance Factors Analysis (PFA) model [1]. Although the PFA-Categorization model tracks the influence of similarities between temporally neighboring items and can be used to predict learners' performance, the model still does not account for the spacing effect.

There are some models solely addressing the spacing effect without considering attentional factors [8, 23, 29]. Among these, the Logistic Knowledge Tracing (LKT) framework developed by Pavlik and colleagues [23] allows testing various features addressing the spacing effect, as detailed in the methods section. Therefore, in this study, we concentrate on utilizing LKT to construct new computational models that integrate both attentional factors and spacing features for predicting student performance in categorical learning with different training sequences (interleaving and blocking). By infusing cognitive features into computational modeling, this approach enhances the model's ability to implement and track human learning, leading to more accurate predictions of learning outcomes.

## 2. METHOD
## 2.1 Spacing in the LKT models

Logistic regression models have been widely used in tracing learners' learning and predicting their retention for their flexibility in incorporating various factors that impact learners' memory processes. The LKT framework facilitates the comparison of various student models, including AFM, PFA, recent-PFA (R-PFA), and others with various novel features [23]. The primary distinction between AFM and PFA lies in their approach to tracking past experiences: PFA tracks counts of prior successes and failures separately, while AFM focuses on counts of prior practices regardless of correctness. In the LKT framework, the efficiency of constructing novel logistic models with tailored features is facilitated through adherence to a symbolic notation system. For example, a $ indicates that a feature fits with one coefficient for each knowledge component (e.g., KC, learner, or item) level, while without a $ sign, all levels of a component are assumed to be treated the same. More details regarding the symbolic notation can be found in Pavlik and Eglington's LKT R package paper [22]. To capture the temporal spacing within sequences, multiple features can be employed within LKT, including *ppe*, *base4*, and *recency*.

The *ppe* feature is the predictive performance equation described in [29], and the *base4* was initially proposed by Pavlik et al. [23] as a comparison to the *ppe* feature. Both features are rooted in the ACT-R memory model [21] and require four parameters to scale spacing between trials and characterize forgetting. A power-law decay is employed to represent the forgetting of successive presentations of items. A primary difference between *ppe* and *base4* lies in their approaches to estimating forgetting rates. Specifically, *ppe* uses all past trials' ages since past practices for its estimation, while *base4* considers only the age of the first trial, representing the time elapsed since the creation of the memory trace. Consequently, the *ppe* feature may be more accurate at tracing spacing and decay. Specifically, as the spacing between consecutive presentations of an item increases, the decay rate for that item decreases. In contrast, the *base4* feature characterizes the spacing between trials by multiplying the mean spacing to a fractional power so it is simpler but less sensitive. When the fractional power falls between 0 and 1, there are diminishing marginal returns for increasing average spacing between trials. When the fractional power is equal to or close to 0, the scaling factor for spacing is 1.

The *recency* feature was developed to capture the recency effect, which suggests that learners tend to perform improved memory retention and recall for items presented to them most recently. The *recency* feature requires one parameter to characterize the impact of the recency of the preceding repetition only, with its value computed as $t^{-d}$, where $t$ is the time elapsed since the prior repetition at the time of the new prediction and $d$ captures the nonlinear decay. While *recency* is not a direct measure of spacing, it can capture the spacing caused performance difference between interleaving and blocking. For instance, in a blocked sequence where the same category is practiced consecutively, there is little decay due to minimal elapsed time between repetitions. Conversely, in an interleaved sequence where items are dispersed, the longer intervals lead to more substantial decay. This approach allows the *recency* feature to account for the immediate effects of spacing, demonstrating its adaptability in modeling cognitive processes related to memory performance.

Table 1. Example models with/no spacing feature in LKT

| Models | Predictors: *feature(component)* |
|---|---|
| AFM | Logit$dec$ (𝐴𝑛𝑜𝑛. 𝑆𝑡𝑢𝑑𝑒𝑛𝑡. 𝐼𝑑) + *intercept*(Problem.Name) + line*afm* (𝐾𝐶. . 𝐷𝑒𝑓𝑎𝑢𝑙𝑡.) |
| PFA | Logit$dec$ (𝐴𝑛𝑜𝑛. 𝑆𝑡𝑢𝑑𝑒𝑛𝑡. 𝐼𝑑) + *intercept*(Problem.Name) + line*suc* (𝐾𝐶. . 𝐷𝑒𝑓𝑎𝑢𝑙𝑡.) + line*fail* (𝐾𝐶. . 𝐷𝑒𝑓𝑎𝑢𝑙𝑡.) |
| PFA + recency | Logit$dec$ (𝐴𝑛𝑜𝑛. 𝑆𝑡𝑢𝑑𝑒𝑛𝑡. 𝐼𝑑) + *intercept*(Problem.Name) + line*suc* (𝐾𝐶. . 𝐷𝑒𝑓𝑎𝑢𝑙𝑡.) + line*fail* (𝐾𝐶. . 𝐷𝑒𝑓𝑎𝑢𝑙𝑡.) + recency (𝐾𝐶. . 𝐷𝑒𝑓𝑎𝑢𝑙𝑡.) |
| PFA + ppe | Logit$dec$ (𝐴𝑛𝑜𝑛. 𝑆𝑡𝑢𝑑𝑒𝑛𝑡. 𝐼𝑑) + *intercept*(Problem.Name) + line*suc* (𝐾𝐶. . 𝐷𝑒𝑓𝑎𝑢𝑙𝑡.) + line*fail* (𝐾𝐶. . 𝐷𝑒𝑓𝑎𝑢𝑙𝑡.) + ppe (𝐾𝐶. . 𝐷𝑒𝑓𝑎𝑢𝑙𝑡.) |
| PFA + base4 | Logit$dec$ (𝐴𝑛𝑜𝑛. 𝑆𝑡𝑢𝑑𝑒𝑛𝑡. 𝐼𝑑) + *intercept*(Problem.Name) + line*suc* (𝐾𝐶. . 𝐷𝑒𝑓𝑎𝑢𝑙𝑡.) + line*fail* (𝐾𝐶. . 𝐷𝑒𝑓𝑎𝑢𝑙𝑡.) + base4 (𝐾𝐶. . 𝐷𝑒𝑓𝑎𝑢𝑙𝑡.) |

We initially examined the impact of spacing in student models by incorporating features such as *ppe*, *base4*, and *recency*, (see example models in Table 1). Given that the AFM is the most basic logistic regression model, we start with AFM. Additionally, we explored the PFA model, leading us to evaluate both AFM and PFA with spacing features. In Table 1, each predictor is presented by the feature (component) format. Logit$dec$ (𝐴𝑛𝑜𝑛. 𝑆𝑡𝑢𝑑𝑒𝑛𝑡. 𝐼𝑑) a single coefficient to characterize the exponential decay of the logit of prior success divided by failures for the participant, serving to capture student variability. *intercept* (*Problem.Name*) represents the initial difficulty of items. *Lineafm* ($KC..Default.$) is the AFM, predicting performance as a linear function of prior practices with the knowledge component. *Linesuc* and *Linefail* are from the PFA model, tracking counts of prior successes and failures.

## 2.2 Attentional Factors

In the second step, attentional features were integrated into the models. Drawing from the sequential attention hypothesis, which suggests that interleaving directs attention to differences between categories, while blocking focuses on similarities within categories [4, 5], both types of comparisons play a crucial role in category learning. In the blocked sequence, there are more within-category comparisons, as all exemplars from the same category are learned consecutively. In contrast, in the interleaved sequence, more between-category comparisons occur, as each exemplar is followed by one from a different category.

Tracking whether consecutive trials belong to the same or different categories can reveal whether participants are engaging in more between-category or within-category comparisons, providing insight into whether the training involves blocking or interleaving. Therefore, to integrate attentional factors, we recorded the counts of comparisons with each preceding trial, considering whether it belongs to the same or different category (see example models with attentional factors in Table 2). We denoted models with attentional factors by adding an "a" before the traditional AFM and PFA models. Specifically, $lineafm$ ($KC..Default.$ % Comparison% Same) tracks the counts of within-category comparisons for each KC and $lineafm$ ($KC..Default.$ % Comparison% Different) tracks the counts of between-category comparisons for each KC.

**Table 2. Example models with attentional factors in LKT**

| Models | Predictors: *feature(component)* |
|---|---|
| a-AFM | Logit$dec$ ($Anon.Student.Id$) + $intercept$(Problem.Name) + lineafm ($KC..Default.$ % Comparison% Same) + lineafm ($KC..Default.$ % Comparison% Different) |
| a-PFA | Logit$dec$ ($Anon.Student.Id$) + $intercept$(Problem.Name) + linesuc ($KC..Default.$ % Comparison% Same) + linefail ($KC..Default.$ % Comparison% Same) + linesuc ($KC..Default.$ % Comparison% Different) + linefail ($KC..Default.$ % Comparison% Different) |
| a-AFM +recency | Logit$dec$ ($Anon.Student.Id$) + $intercept$(Problem.Name) + lineafm ($KC..Default.$ % Comparison% Same) + lineafm ($KC..Default.$ % Comparison% Different) + recency ($KC..Default.$) |

## 2.3 Cross Validation

In this study, a student-stratified cross-validation approach was employed to rigorously validate the entire process, ensuring no leakage of model information between different iterations for the held-out data. The dataset was randomly partitioned into five folds, with models trained on four folds and tested on the fifth. This cross-validation procedure was iterated ten times for each model, and the means of key model fit parameters, including $R^2$, AUC, RMSE, and correlation coefficients, were calculated for the test fold. The correlation coefficients, in particular, played a crucial role in assessing the model's ability to predict both learning session and posttest performance accurately. The reported averages represent a comprehensive evaluation of the model's predictive capabilities across multiple iterations, providing a robust measure of its overall performance.

## 3. DATASETS

### 3.1 Bird Category Learning

The bird learning dataset included 43,326 observations from 181 participants who learned bird categories by reviewing bird images and selecting which category it is, after excluding outliers. The study was approved by the Institutional Review Board of the University of Memphis. Participants were Amazon Mechanical Turk workers recruited through web service (www.cloudresearch.com). Participants were randomly assigned to four training groups: high similarity narrow spacing, high similarity wide spacing, low similarity narrow spacing, and low similarity wide spacing. All participants completed a pretest consisting of 40 items (bird photos: 10 categories × 4 exemplars to be learned). Subsequently, participants engaged in a learning session with five levels of block size: 1 (fully interleaving), 2, 4, 8, and 16 (fully blocking). The stimuli were repeated 4 times (160 trials: 10 categories × 4 exemplars × 4 repetitions). Therefore, each learning session comprises a total of 168 trials. Finally, participants completed a posttest featuring 60 items (40 learned items and 20 novel stimuli). Participants received feedback during the learning session but not during the pretest and posttest. The stimuli type, the manipulation of block size and spacing, repetition of KCs and items made this dataset appropriate for evaluating model fit to the effects of training sequences (interleaving vs. blocking). The data will be available by request.

### 3.2 Blob Figures Learning

In Carvalho and Goldstone's [3] first experiment, they investigated the impact of category structure (high vs. low similarity) on the efficacy of interleaving and blocking. The dataset comprises 40,320 observations from 60 participants after excluding outliers. Participants were undergraduate students from Indiana University. They were randomly assigned to either the high similarity group or the low similarity group. All participants initiated the study by learning three categories in either interleaved or blocked sequences, followed by testing. Subsequently, they underwent another study phase in a different sequence and were tested again. Each study phase consisted of four sessions, with 72 trials in each session. Carvalho and Goldstone [3] observed that interleaved sequence enhanced the classification of novel items at the posttest for high similarity categories, whereas blocked sequence improved the classification of novel items at the posttest for low similarity categories. The manipulation of training sequences in this dataset also rendered it suitable for assessing model fit concerning the effects of training sequences. The data can be found at https://osf.io/t8wzy/

## 4. RESULTS

### 4.1 Bird Category Learning

See Table 3 and Figure 1 for the model fitting results of AFM and PFA models incorporating spacing features. Notably, before adding spacing features, there was no discernible difference between the original AFM and PFA models concerning the $R^2$, AUC, and RMSE. Both models demonstrated proficient predictions for the learning session. However, on the posttest, both AFM and PFA models demonstrated poor predictive performance, with the PFA model showing no correlation with the human data.

**Table 3. Mean test fold fitting results with/no spacing for bird data**

| Model | $R^2$ | AUC | RMSE | $r_1$ | $r_2$ |
|---|---|---|---|---|---|
| AFM | 0.21 | 0.80 | 0.40 | 0.87 | 0.24 |
| PFA | 0.21 | 0.80 | 0.40 | 0.90 | -0.05 |
| AFM + recency | 0.29 | 0.85 | 0.38 | 0.98 | 0.31 |
| PFA + recency | 0.29 | 0.85 | 0.38 | 0.98 | -0.18 |
| AFM + ppe | 0.30 | 0.85 | 0.38 | 0.99 | 0.44 |
| PFA + ppe | 0.31 | 0.86 | 0.38 | 0.99 | -0.24 |
| AFM + base4 | 0.23 | 0.82 | 0.40 | 0.87 | 0.29 |

*Note.* $r_1$ represents the correlation between the model prediction and the human data for the learning session. $r_2$ represents the correlation for the posttest.

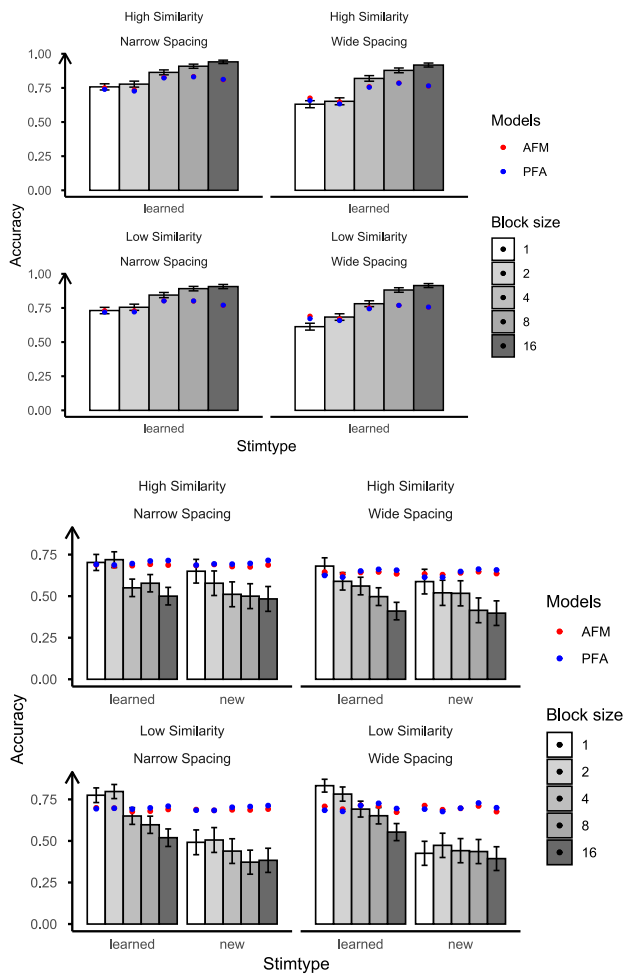

**Figure 1.** Fitting results (dots) of the learning session (top panel) and posttest (bottom panel) for the AFM and PFA model over the empirical results from bird data (represented by the bars).

Upon the inclusion of spacing features, both *recency* and *ppe* exhibited an enhancement in model fit, while *base4* did not contribute significantly to the improvement. Therefore, in subsequent analyses, we focused solely on testing the effects of *recency* and *ppe*.

For the AFM, the addition of spacing features led to improved predictive performance for both the learning session and the posttest. However, for the PFA models, the inclusion of spacing features resulted in more accurate predictions for learning sessions but a decline in posttest predictions. During the learning session, the larger the block size, the higher the accuracy of participants' bird categorization. However, the posttest showed a reverse pattern that the smaller block size led to higher accuracy. This discrepancy challenges the predictions of the PFA model, which tracks counts of prior successes and failures. While the PFA model suggests that more correct answers indicate more learning, this assumption does not hold true in the context of interleaving or blocking. In a blocked sequence, where items from the same category are practiced consecutively, participants may produce correct responses more easily. Yet, this doesn't necessarily imply effective learning. Consequently, the PFA models may overfit the learning session data and exhibit bias against the posttest results. Therefore, in the following section of testing the attentional factors, we only explored the AFM.

**Table 4. Mean test fold fitting results with attentional factors for bird data**

| Model | $R^2$ | AUC | RMSE | $r_1$ | $r_2$ |
|---|---|---|---|---|---|
| a-AFM | 0.21 | 0.80 | 0.40 | 0.98 | 0.56 |
| a-AFM+recency | 0.29 | 0.85 | 0.38 | 0.98 | 0.55 |
| a-AFM+ppe | 0.31 | 0.85 | 0.38 | 0.98 | 0.79 |

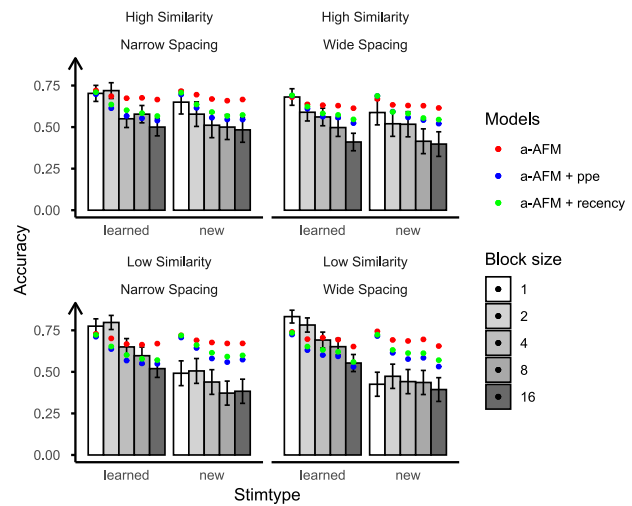

**Figure 2.** Fitting results (dots) of the posttest (bottom panel) for the attention-based AFM models over the empirical results from bird data (represented by the bars).

We initiated the examination by testing the AFM with attentional factors, specifically focusing on the counts of between-category comparisons and within-category comparisons. The outcomes are presented in Table 4 and Figure 2. Utilizing attentional factors in isolation did not enhance overall model fit significantly but did result in improved predictions for both the learning session and posttest (see $r_1$ and $r_2$ in Table 4). Subsequently, we introduced spacing into the attention-based AFM. With the integration of both spacing and attentional factors into the model, there was an overall enhancement in model fit. However, *recency* and *ppe* demonstrated distinct effects on model predictions. The addition of *recency* to the attention-based AFM did not enhance the model's predictive accuracy for the posttest. In contrast, incorporating *ppe* resulted in an improvement in posttest predictions. This difference may be attributed to *recency* having only one parameter, whereas *ppe*, with four parameters, could better capture practice curves, spacing, and forgetting. In summary, both spacing and attentional factors are crucial in capturing the influence of interleaving and blocking on category learning. It's worth noting that the bird experiment design includes two between-subject variables: similarity and spacing, along with a within-subject variable, training sequence. However, the current model primarily focuses on the training sequence and spacing without incorporating the similarity parameters. The distinctions between high and low similarity, as depicted in Figure 2, were captured by $Logitdec(Anon.Student.Id)$. Although adding the similarity feature falls beyond the scope of the current paper, future research could explore its inclusion to enhance model fit.

## 4.2 Blob Figures Learning

**Table 5. Mean test fold results for blob figures learning data**

| Model | $R^2$ | AUC | RMSE | $r_1$ | $r_2$ | $r_3$ |
|---|---|---|---|---|---|---|
| AFM | 0.14 | 0.75 | 0.43 | 0.99 | 0.92 | 0.36 |
| PFA | 0.15 | 0.76 | 0.42 | 0.98 | 0.85 | 0.02 |
| AFM+recency | 0.16 | 0.76 | 0.42 | 0.99 | 0.94 | 0.51 |
| a-AFM+recency | 0.16 | 0.76 | 0.42 | 0.99 | 0.95 | 0.65 |
| AFM+ppe | 0.15 | 0.76 | 0.42 | 0.99 | 0.96 | 0.68 |
| a-AFM+ppe | 0.15 | 0.76 | 0.42 | 0.99 | 0.96 | 0.73 |

*Note.* $r_1$ denotes the correlation between the model prediction and human data for the learning session. $r_2$ represents the overall correlation between the model prediction and human data for the posttest. $r_3$ specifically signifies the correlation on the posttest for high-similarity stimuli.

To assess the replicability of our findings from the bird category learning data, we conducted an evaluation of model fits and predictions using the blob figures learning dataset. All models demonstrated accurate predictions during the learning sessions (see Table 5). However, when examining the posttest which includes learned and new items, we found that the posttest data exhibited distinct patterns based on the similarity of stimuli (see Figure 3). In instances of low similarity new items, participants demonstrated enhanced generalization for blocked items, mirroring the learning session. Conversely, for high-similarity stimuli, participants exhibited a reversed pattern, showing improved performance on interleaved items, contrary to the learning session. The AFM model displayed relatively accurate predictions for low similarity but faltered for high similarity stimuli (see $r_3$ value in Table 5 and Figure 3 bottom panel). The incorporation of both *recency/ppe* and attentional factors into the model resulted in improved predictions for high-similarity stimuli. Specifically, the model integrating both attentional factors and *ppe* showed the most improvement in posttest predictions. This observation aligns with the findings from the bird experiment. Generally, the findings underscored the importance of both spacing features and attentional factors in capturing the effects of training sequences (interleaving vs. blocking).

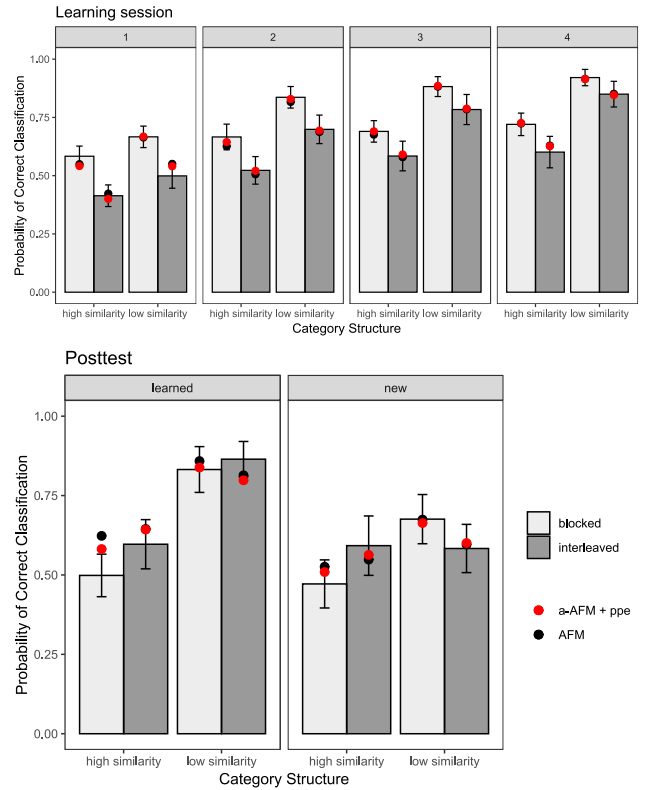

**Figure 3. Fitted model results (dots) of the learning session (top panel) and posttest (bottom panel) for the AFM and a-AFM+ppe models for the blob figures learning data (represented by the bars).**

## 5. DISCUSSIONS

The present study incorporated spacing features and attentional factors into LKT models to examine the influence of training sequence (interleaving vs. blocking) on category learning. Through cross-validation, we evaluated the model fit and examined the correlations between model predictions and human data during both learning sessions and posttests. Results from model fitting analyses on two datasets emphasize the crucial role of both spacing and attentional factors in understanding the effects of training sequences (interleaving and blocking) on category learning. Notably, instances where the training sequence results in a reversed performance trend between the learning session and posttest highlight the importance of both attentional factors and the spacing features. The finding aligns with and supports the validity of the two-stage framework [31].

Our findings suggest that basic logistic knowledge tracing models, including AFM and PFA, effectively predict learners' category learning performance during learning sessions but encounter challenges when forecasting posttest performance. This aligns with research examining the efficacy of Bayesian Knowledge Tracing (BKT) and the AFM in monitoring student learning and predicting future performance [13, 25, 27]. Researchers found that despite AFM, BKT, and BKT-F (BKT with forgetting) have the ability to capture qualitative learning trends across sessions and achieve acceptable fit metrics, all models fall short in capturing the spacing effect over multiple sessions [27]. Surprisingly, the BKT-F, which has a mechanism accounting for forgetting and thus should predict the benefits of spacing failed to make accurate predictions on subsequent performance after the manipulation of spacing intervals during initial learning.

After incorporating spacing features (*recency* and *ppe*) into both the AFM and PFA models, there was an improvement in model fit for bird category learning data, while the improvement was not as pronounced in the blob figures learning data regarding the $R^2$, AUC, and RMSE values. This discrepancy may be attributed to differences in experimental design. In the bird category learning experiment, which involved 10 bird categories, the manipulation of spacing intervals (wide vs. narrow) may have provided more variability in the data, allowing the spacing features to have a more pronounced effect on model fit. Whereas the blob figures learning experiment, with only 3 categories in each sequence resulted in unintentionally narrower spacing intervals, may have had less variability in the spacing effect, thus resulting in a less significant improvement in model fit.

A shared advantage of integrating spacing features into the AFM across both datasets is the enhanced correlation between model predictions and human data, particularly in posttest predictions. This improvement is especially noticeable when a reverse performance pattern occurs between the learning session and the posttest. The rationale behind this lies in the nature of interleaved and blocked sequences. In blocking, performance during the learning session may be artificially inflated due to the ease of guessing correctly (lack of spacing). In contrast, interleaving sequences, despite potential challenges during the learning session, benefit from the spacing effect, resulting in superior posttest performance compared to blocked items. Therefore, the inclusion of spacing features helps capture these differences in spacing between the two training sequences, leading to more accurate performance predictions on the posttest.

Although the inclusion of attentional factors did not boost the overall model fit significantly, it did contribute to enhanced predictions of the posttest fit. This enhancement stems from the role of attentional factors, which use counts of prior comparisons (whether same or different) to indicate the training sequence. Notably, interleaving involves more between-category comparisons, while blocking entails more within-category comparisons. Thus, both spacing and attentional factors play a crucial role in capturing learners' performance across different sequences (interleaving and blocking) in category learning.

In future research, there is a need to expand the application of attentional factors and spacing within AFM models to a broader range of datasets, ensuring the model's efficacy across diverse educational contexts. Besides, conducting comparative analyses with other knowledge tracing models, such as the BKT and BKT-F, is imperative to validate the distinctive contributions of the new model and identify potential areas for further enhancement. Additionally, the current datasets only feature short retention intervals, with the posttest immediately following the learning session. Future investigations should explore the adaptability of the model to longer retention intervals and diverse learning sequences, thereby enhancing the model's practical utility.

## 6. ACKNOWLEDGMENTS

Our thanks to Paulo Carvalho for providing the blob figures learning data analyzed in this study.